\def\two_column{0}

\if\two_column1
\documentclass[10pt, twocolumn,a4paper,final]{article}
\def\gscale{1.0}
\else
\documentclass[10pt, a4paper,final]{article}
\def\gscale{0.6}
\fi
\usepackage[numbers,sort&compress]{natbib}
\usepackage{amsmath}
\usepackage{amssymb}
\usepackage{bm}
\usepackage[hyphens]{url}
\usepackage{graphicx,epstopdf}% Include figure files
\usepackage{paralist}
\usepackage[margin=2cm]{geometry}
\usepackage[font=small,labelfont=bf]{caption}
\usepackage {color}
\usepackage{pdfpages}

\newcommand{\ignore}[1]{}

\begin{document}

\title{\bf
%Ionic Coulomb Blockade,
Quantized Dehydration
and the Determinants
of Selectivity in the NaChBac Bacterial Sodium Channel
}
\author{
O. A. Fedorenko$^1$,
I. Kh. Kaufman$^2$,
W. A. T. Gibby$^2$,
D. G. Luchinsky$^{2,3}$\\
%C. Guardiani,
%I. A. Khovanov,
S. K. Roberts$^1$
P. V. E. McClintock$^2$ \\
{\footnotesize $^1$Division of Biomedical and Life Sciences, Lancaster University, Lancaster, LA1 4YQ, UK}\\
{\footnotesize $^2$Department of Physics, Lancaster University, Lancaster, LA1 4YB, UK}\\
{\footnotesize $^3$SGT, Inc., Greenbelt, MD, 20770, USA}
}

\maketitle

\begin{abstract}

\noindent A discrete electrostatic/diffusion model has been developed to describe the selective permeation of ion channels, based on ionic Coulomb blockade (ICB) and quantised dehydration (QD). It has been applied to describe selectivity phenomena measured in the bacterial NaChBac sodium channel and some of its mutants. Site-directed mutagenesis and the whole-cell patch-clamp technique were used to investigate how the value  $Q_f$ of the fixed charge at the selectivity filter (SF) affected both valence and alike-charge selectivity. The new ICB/QD model predicts that increasing ${Q_f}$ should lead to a shift of selectivity sequences towards larger ion sizes and charges, a result that agrees with the present experiments and with earlier work. Comparison of the model with experimental data provides evidence for an {\it effective charge} $Q_f^*$ at the SF that is smaller in magnitude than the nominal $Q_f$ corresponding to the charge on the isolated protein residues. Furthermore, $Q_f^*$ was different for aspartate and glutamate charged rings and also depended on their  position within the SF. It is suggested that protonation of the residues within the restricted space is an important factor in significantly reducing the effective charge of the EEEE ring. Values of $Q_f^*$ derived from experiments on the anomalous mole fraction effect (AMFE) agree well with expectations based on the ICB/QD model and have led to the first clear demonstration of the expected ICB oscillations in Ca$^{2+}$ conduction as a function of the fixed charge. Pilot studies of the dependence of Ca$^{2+}$ conduction on pH are consistent with the  predictions of the model.

\end{abstract}
%\subsection*{Keywords}
%Fixed charge, quantized dehydration, selectivity, selectivity filter, ionic Coulomb blockade, bacterial channel, protonation, site-directed mutagenesis
%\tableofcontents

\section*{Introduction}
\label{sec:intro}
Biological ion channels provide for the highly-selective passive transport of physiologically important ions (e.g.\ Na$^+$, K$^+$ and Ca$^{2+}$) through the bilipid membranes of living cells. The channels consist of nanopores within complex proteins embedded in the membrane. Their selectivity for particular cations is determined by the electrostatically-driven stochastic motion of ions within a short and narrow  selectivity filter (SF) carrying a binding site with fixed negative charge $Q_f$ \cite {Hille:01}.
Following Eisenman \cite {Eisenman:83} ionic selectivity arises through a balance between repulsion by the dehydration/self-energy barrier and electrostatic attraction/affinity to the  binding site, resulting in {\it resonant barrier-less conduction} for the selected ion \cite{Laio:99,Berneche:01,Nadler:03,Allen:04,Yesylevskyy:05,Corry:17}, and leading to selectivity phenomena such as the divalent blockade of the sodium current \cite{Woodhull:73,Sather:03} and the anomalous mole fraction effect (AMFE) \cite{Gillespie:08b,Sather:03}.

The resonant barrier-less permeation can be described in terms of ionic Coulomb blockade (ICB), an electrostatic phenomenon that appears in low-capacitance mesoscopic systems
\cite{Kitzing1992,Zwolak:09,Krems:13,Kaufman:13a,Kaufman:13b,Kaufman:15,Kaufman:15b}. ICB predicts $Q_f$ to be an important determinant of selectivity, and one that is manifested strongly for divalent ions e.g.\ by giving rise to Ca$^{2+}$ conduction bands \cite {Kaufman:13a}. ICB is closely similar to its electronic counterpart in quantum dots \cite{Averin:86,Beenakker:91} and nanostructures\cite{Grabert:13,Feng:16}.
The basic ICB model for the permeation and selectivity of ion channels \cite{Kaufman:15b} has recently been enhanced  \cite{Kaufman:17b, Luchinsky:17a} by the introduction of shift/corrections to allow for the singular part of the ionic attraction to the binding site (i.e.\ local site-binding), as well as for arbitrary additional ion excess chemical potential $\Delta \mu_{ex}$. The geometry-dependent shift of the ICB calcium resonant point, following from this correction, leads to a change in the divalent (calcium) blockade threshold $IC_{50}$ \cite{Kaufman:17b}.

The voltage-gated bacterial sodium channels NaChBac, NavAb, NavMs, and NvsBa are a family of relatively simple channels with discovered structures. Hence they are widely used in modelling the general features of conductivity and selectivity \cite{Ren:01,Yue:02,DeCaen:14,Finol:14,Naylor:16,Guardiani:17,Catterall:17,Payandeh:15}.
Site-directed mutagenesis, varying the fixed charge $Q_f$ at the SF, is known to change their selectivity, switching sodium channels to calcium and {\it vice versa}. In turn, the alteration of similarly-charged glutamate residues to aspartate was also found to influence the channels' conductivity and selectivity \cite{Yue:02,DeCaen:14,Finol:14,Naylor:16}. The nature and physical origin of such transformations has remained unclear.

In this paper, we further enhance the ICB model by inclusion of the effect of quantised dehydration (QD) \cite {Laio:99,Zwolak:09,Corry:17}. The resultant ICB/QD model ia applied to an analytic, numerical and experimental study of the effect of the fixed charge $Q_f$ on the conductance and selectivity of NaChBac bacterial sodium channels and relevant mutants \cite {Yue:02,Naylor:16,Guardiani:17,Kaufman:17a,Guardiani:17b}. In doing so, we develop a novel picture of {\it resonant quantized dehydration}, combining the idea of quantized (shell-based) dehydration with the balance/shift-enhanced ICB model. Our systematic mutation study of selectivity in the NaChBac channel will be used to show that the ICB/QT model can account for the experimental Eisenman sequences and for measurements of the anomalous mole fraction effect (AMFE) in the mutants.

In what follows, with SI units, $\varepsilon_0$ is the permittivity of free space, $e$ is the elementary charge,  $k_B$ is Boltzmann's constant and $T$ is the temperature. We use the conventional shorthand symbols for amino acid residues: L--Leucine; E--Glutamate ($Q_f=-1$); D--Aspartate ($Q_f=-1$);  S--Serine; W--Tryptophan; A--Alanine; K--Lysine ($Q_f=+1$);  T--Threonine; and so on.

%-----------------------------------------
\begin{table}
	\begin{center}
		\resizebox{\gscale\columnwidth}{!}{%
			\begin{tabular}{|c| c c c c c c|c|}
				\hline
				Mutant   &  \multicolumn{6}{c|}{Selectivity Filter Sequence} & $Q_f^{nm}/e$\\
				channels & 190 &\color{blue}{191}&\color{blue}{192} &193 & 194 & 195 &  \\
				\hline
				Wild type& L & \color{blue}{E} &\color{blue}{S } & W & A & S  & -4  \\
				%	\hline
				E191D&L & \color{red}{D} & \color{blue}{S} &W  &A  & S  & -4\\
				%	\hline
				&  &  &  &  &  &  &\\
				%\hline
				E191A & L & \color{red}{A} & \color{blue}{S} & W & A & S & 0  \\
				%	\hline
				S192K & L & \color{blue}{E} & \color{red}{K} & W & A &S & 0 \\
				%	\hline
				&  &  &  &  &  & &\\
				%	\hline
				S192E & L &\color{blue}{E}  &\color{red}{E}  &W  & A & S & -8 \\
				%	\hline
				S192D & L & \color{blue}{E} &\color{red}{ D} & W & A & S & -8 \\
				%\hline
				E191D, S192E	& L & \color{red}{D} & \color{red}{E} & W & A & S & -8 \\
				%\hline
				E191D, S192D	& L &\color{red}{D}  &\color{red}{D}  & W &A  &S  & -8\\
				\hline
			\end{tabular}
		}
	\end{center}
	\caption{The wild-type NaChBac channel and its mutants studied in this papar, showing the amino acid sequences in their SFs and the corresponding nominal values of fixed charge $Q_f^{nm}$. The key SF positions (191,192) are shown in blue, except for changed residues which are highlighted in red. The mutants are grouped by their $Q_f^{nm}$ values. Note that for the S192K mutant the neutral serine residue is replaced by a positively charged lysine which is expected to neutralise the negatively-charged sidechain og the glutamate resulting in a $Q_f^{nm}$ value of 0. %\footnotesize{Shorthand symbols used for amino acid residues are: L -  Leucine, E - Glutamate ($Q_f^{nm}$=-1e), D - Aspartate ($Q_f^{nm}$=-1e),  S - Serine, W - Tryptophan, A- Alanine, K - Lysine ($Q_f^{nm}$=+1e)
	}
	\label{tab:mutants}
\end{table}
%------------------

\section*{Experimental materials and methods}

\subsection*{Channels/mutants studied}
The voltage-gated NaChBac bacterial channel \cite{Ren:01,Yue:02} is a tetrameric channel, whose SF is formed by 4 trans-membrane segments each containing the six-amino-acid sequence LESWAS, corresponding to residues 190 –- 195). This structure provides the highly-conserved \{EEEE\} locus E191 with a nominal $Q_f=-4e$ which is considered to create a single binding site for both mono- and  divalent moving ions \cite {Yue:02,Ren:01}. Table \ref{tab:mutants} presents the set of channels generated and studied in the current research.

\subsection*{Generation and expression of wild-type and mutant NaChBac channels}

The NaChBac (GenBank accession number BAB05220) cDNA construct containing 274 amino acid residues was synthesised by EPOCH Life Science (www.epochlifescience.com) and subcloned into the mammalian cell expression vector pTracer-CMV2 (Invitrogen). Single amino acid mutations in the pore region of NaChBac were generated by site-directed mutagenesis using oligonucleotides containing the sequence for the desired amino acid substitutions (primers are listed in supplemental table 1) and Q5 Site-Directed Mutagenesis Kit (New England BioLabs Inc.). All mutations were confirmed by DNA sequencing. Wild-type (wt) NaChBac and mutant cDNAs were transiently transfected into CHO cells with TransIT-2020 (Mirus Bio). Transfected cells were identified by GFP fluorescence using an inverted fluorescence microscope (Nikon TE2000-s) and used for electrophysiological investigation 24 –-48 hours after transfection.

\subsection*{Electrophysiology}
Whole-cell voltage clamp recordings were performed at room temperature (20$^\circ$C) using an Axopatch 200A (Molecular Devices, Inc.) amplifier. Whole cell currents were elicited by a series of step depolarizations (+95mV to -85mV in -15mV steps) from $V_{hold}$ of -100mV.
Details of the patch-clamp methods are presented in the Supplemental Information.

\section*{Results and discussion. %Mutation study of selectivity in NaChBac channels
}
\label{sec:exp}
The main aims of the research were to establish the dependence of the conductivity and selectivity of NaChBac  mutants on the fixed charge $Q_f$ at the SF, and also on (D$\rightleftharpoons$E) substitutions within the SF, in order to see whether the results could be understood within the framework of the ICB model.  Increasing the value of $Q_f$ was expected to lead to stronger divalent blockade following the Langmuir isotherm and to a resonant variation of the divalent current with $Q_f$ \cite{Kaufman:15,Kaufman:17a}.

Previous mutant studies \cite{Yue:02,Ren:01,DeCaen:14,Naylor:16} investigated a limited number of possible (D,E) combinations in the key positions 191 and 192. Our systematic study of the possible mutants L\{ES/DS/EE/ED/DE/DD\}WAS enables us to identify the influence both of the fixed charge $Q_f$ and of D/E substitutions in positions 191 and 192. Here we present selectivity sequences for monovalent and divalent ions, recorded for the mutants listed above, with nominal fixed charge $Q_f$ from $0e$ to $-8e$, with the permutations of D and E shown in Table. \ref{tab:mutants}. We have also made divalent blockade/AMFE measurements, providing us with the experimental information needed for application of the extended ICB model (see below) incorporating the effect of quantised dehydration.

%(see Appendix), with addition of divalent blockade/AMFE measurements.%starting from measuring mono- and divalent  selectivity sequences and divalent blockade/AMFE properties for wild type NaChBac (LESWAS) and its 191D (LDSWAS) mutant (see Tabl. \ref {tab:mutants} )

\subsection*{Resonant quantized dehydration}
%\section*{Resonant selectivity model}

%\subsection*{Extended ionic Coulomb blockade model and shift-based selectivity}
%\label{sec:icb}
We  consider the stochastic transport of a fully-hydrated Ca$^{2+}$ ion with a charge of $q=2e$. The original ICB model, based on first-principles electrostatics and taking account of charge discreteness and an electrostatic exclusion principle, specifies the parameters for the resonant ($\{n\}\rightarrow\{n+1\}$) conduction points $M_n$ and Coulomb blockade/neutralisation points $Z_n$. These are  defined \cite{Kaufman:15,Kaufman:15b} by the condition for barrier-less motion  $\Delta U_q-\mu_F=0$,
\begin{equation}
Z_n^{CB}=-n q;   \   \
M_n^{CB}=-(n+1/2) q.%\nonumber
\end{equation}
The basic ICB model can be extended by consideration of an added free energy per ion. This excess chemical potential $\Delta \mu_{ex}$ arises from different sources \cite{Roux:04,Krauss:11,Boda:13a}. Inclusion of the ICB barrier-less condition $\Delta \mu_{ex}=0$ leads to a ``shift equation'' \cite {Kaufman:17b,Luchinsky:17a,Kaufman:17a}

\begin{equation}
M_0=M_0^{CB}%+\Delta M_0^{TS}+\Delta M_0^{DH} %
-\left(C_s/q\right) \sum_Y{\Delta \mu^{Y}}
\label {equ:alt_shift}
\end{equation}
where the additional energies $\Delta \mu^Y$ lead to shifts $\Delta M_0$ in the resonant barrier-less $Q_f$ point $M_0$  \cite{Kaufman:15b,Kaufman:15,Kaufman:17a}. The most important of these are:

\begin{compactitem}
\item A concentration-related bulk entropy shift $\Delta \mu^{TS}=-k_BT\log(P_b)$; and
\item A dehydration shift $\Delta \mu^{DH}$.
\end{compactitem}

\noindent Here,  $C_s$ is the electrical self-capacitance of the channel \cite{Kaufman:15} and $M_0^{CB}=-q/2$ is the base ICB  barrier-less point; a similar equation was derived earlier  \cite{Luchinsky:16,Luchinsky:17a}.

%-----------------------------
\begin{figure}[t!]
	\begin{center}
		\includegraphics[width=\gscale\linewidth]{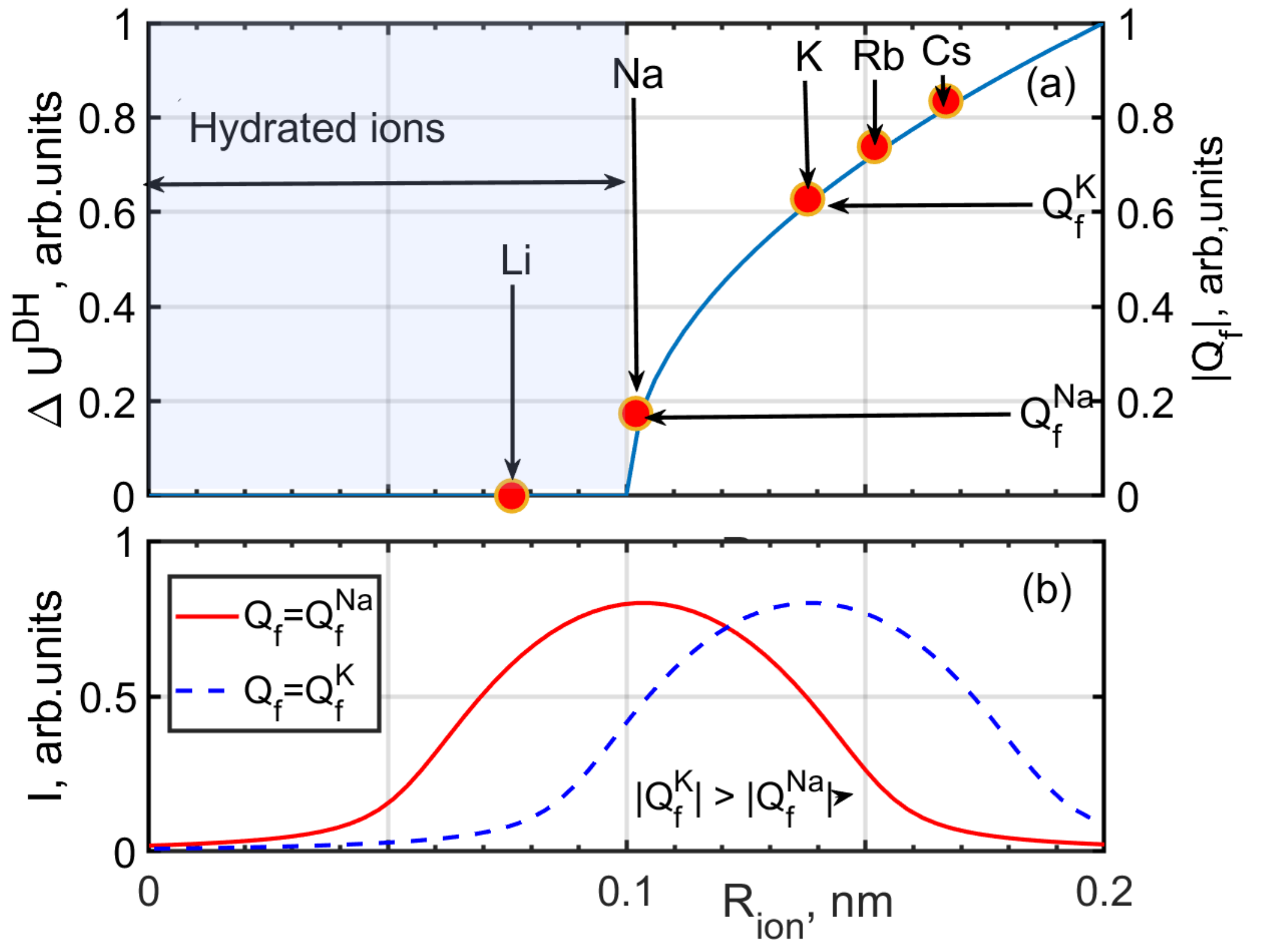}
	\end{center}
	\caption{(Color online). Quantized dehydration scheme for monovalent ions in the  selectivity filter (SF) of the NaChBac channel. The radius of the SF is taken as $R_{c}=0.3$\,nm, and the thickness of the first hydration shell as $h_c=0.2$\,nm.
		(a) The dehydration energy $\Delta U_q^{DH}$ {\it vs.} $R_{ion}$ (blue solid line) is calculated to be proportional to $\sqrt{R_{ion}-(R_c-h_c)}$ . The shaded area indicates the range of radii where ions can fit within the SF while still retaining their 1st hydration shells intact. The right-hand ordinate axis shows the resonant fixed charge corresponding to a given ion's size/dehydration energy.
		(b) Conduction {\it vs.} $R_{ion}$ for different values of the fixed charge $Q_f$.
	}
	\label {fig:DH1}
\end{figure}
%-----------------------------

%------------------
\begin{figure}[t!]
	\begin{center}
		\includegraphics[width=\gscale\linewidth]{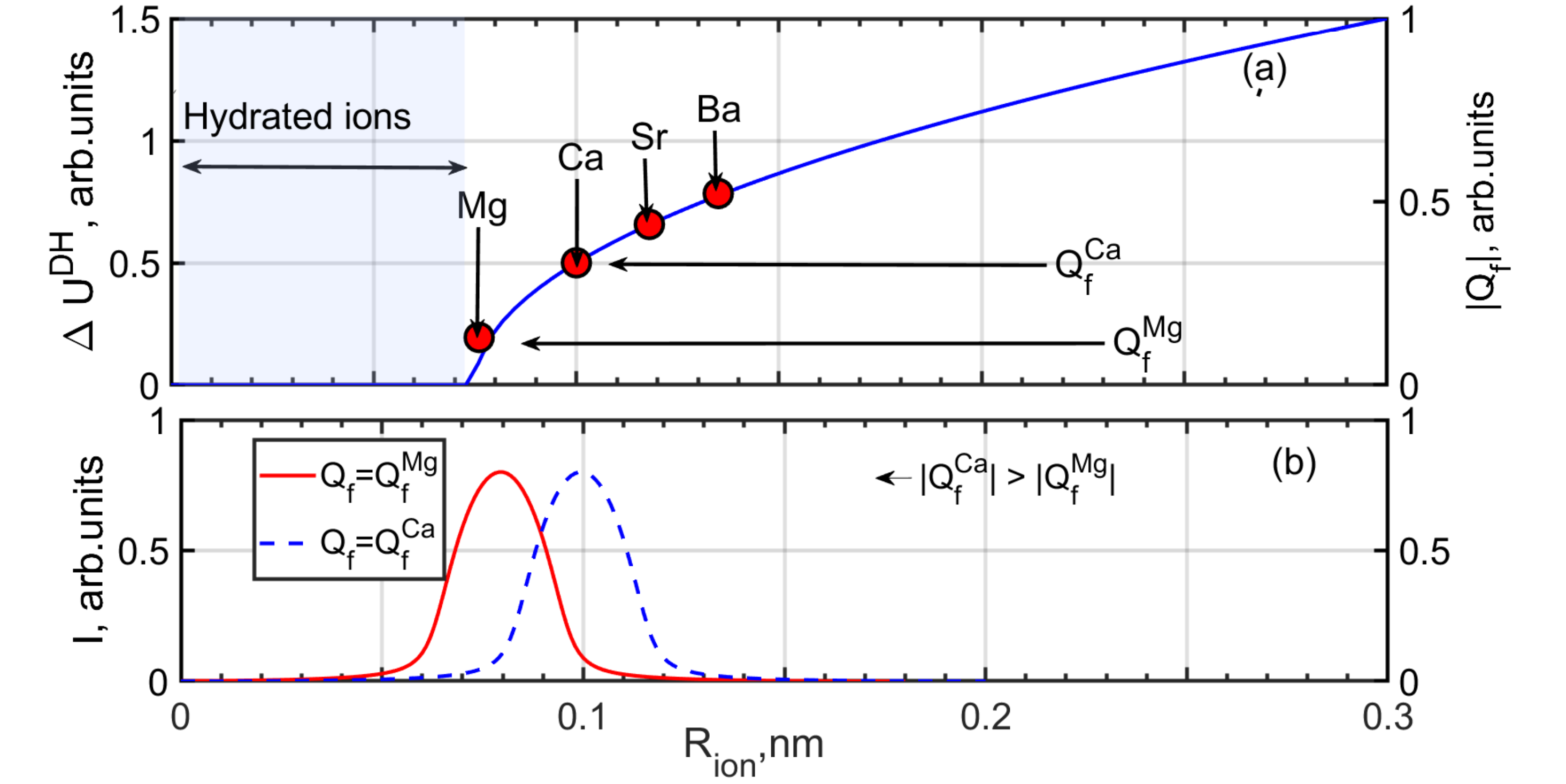}
	\end{center}
	\caption{(Color online). Quantized dehydration scheme for divalent ions in the selectivity filter (SF) of the NaChBac channel. The radius of the SF is taken as $R_{c}=0.3$\,nm, and the thickness of the first hydration shell as $h_c=0.225$\,nm.
		(a) The dehydration energy $\Delta U_q^{DH}$ {\it vs.} $R_{ion}$ (blue solid line) is calculated to be proportional to $\sqrt{R_{ion}-(R_c-h_c)}$ . The shaded area indicates the range of radii where ions can fit within the SF while still retaining their 1st hydration shells intact. The right-hand ordinate axis shows the resonant fixed charge corresponding to a given ion's size/dehydration energy.
		(b) Conduction {\it vs.} $R_{ion}$ for different values of the fixed charge $Q_f$.
	}
	\label {fig:DH2}
\end{figure}
%-----------------------------

Dehydration, full or partial is thought to be the main source of selectivity between equally charged ions, e.g.\ monovalent alkali metal ions \cite{Laio:99,Nonner:98,Sahu:17,Li:17}. The basic ICB model takes account of hydration/dehydration only through the dielectric self-energy $U_q^{SE}$ \cite{Laio:99,Zhang:05} which does not depend on the size of the ion, so additional effects need to be included in the model.

One such effect is the discreteness of the hydration shells, which strongly influences the selectivity \cite{Laio:99,Zwolak:09,Yesylevskyy:05,Corry:17}. Details of the ion-ligand interactions \cite{Corry:12,Corry:17,Dudev:14} and multi-ion knock-on mechanisms \cite {Berneche:01,Hodgkin:55} are also important. A hydrated ion is assumed to be surrounded by spherical, discrete, single-molecule water shells of equal radial thickness $h_c$. The first shell is immediately adjacent to the ion, so the hydrated ion moves as an ion-water complex of radius $R_{ion}^{}=R_{ion}+h_c$. The shell model of hydration has been well-validated by experimental, analytical and numerical evidence. So, consistent with the positions of the minima in experimental and MD-simulated radial density functions \cite{Shao:09,Zwolak:09}, we will take $h=0.2$\,nm for monovalent ions and $h=0.225$\,nm for divalent ions.

We will consider an ion that retains its first shell untouched during its passage through the SF as remaining fully hydrated. Otherwise ions must lose/rearrange their first hydration shells with corresponding energy penalties. While small rearrangements of the shell are relatively cheap energetically, a decrease in the coordination number immediately leads to significant expense.

Generic shell-based quantized dehydration (QD) models provide a simple explanation for the difference in K$^+$/Na$^+$ selectivity between K$^+$ and Na$^+$/Ca$^{2+}$ channels \cite{Laio:99,Corry:17}: within narrow K$^+$ channels, both Na$^+$ and K$^{+}$ ions are fully dehydrated. The first hydration shell is more tightly-bound to the smaller Na$^+$ ion, $\Delta U_{K}^{DH}<\Delta U_{Na}^{DH}$, and the channel therefore favours the larger ion. In contrast, the moderately wide Na$^+$ or Ca$^{2+}$ channels accommodate  both Na$^+$ and K$^+$ ions with their first hydration shells intact. In such cases $\Delta U_{K}^{DH}>\Delta U_{Na}^{DH}$ \cite{Zwolak:09} and the channel favours the smaller Na$^+$ ion. However, generic models of this kind cannot explain the influence of $Q_f$ on the selectivity sequences of a channel.

In order to go further, we now combine the ideas of the QD models \cite{Zwolak:09,Corry:17} with the Eisenman-inspired model of barrier-less selectivity \cite{Eisenman:83,Nadler:03,Yesylevskyy:05} and, in particular, with the enhanced ICB model \cite{Kaufman:15,Kaufman:17b, Luchinsky:17a}. In the picture proposed, the difference in $\Delta U^{DH}$ is not enough {\it per se} to determine which species will be selected over the other as this choice could be changed or even inverted by the value of $Q_f$ needed to provide barrier-less conduction, i.e.\ by the corresponding shift of the resonant point $M_n$. Hence, if $\Delta U^{DH}$ increases with $R_{ion}$ then $|M_n|$ will also increase for the wider Na$^+$/Ca$^{2+}$ channels, but {\it vice versa} decrease for the narrower K$^+$ channels.

We assume that, for an Na$^+$/Ca$^{2+}$ channel ($\epsilon_w\gg \epsilon_p$), the electrostatic field $W$ of the ion inside the channel can be decomposed in terms of a small parameter $\kappa=\epsilon_p/\epsilon_w$ as $W=W^{CB}+W^{DH}=W^{CB}+\kappa W^{SP}$, where:

\begin{compactitem}

\item $W^{CB}$ is the main field, which is 1D (flat) Coulomb field localized inside the channel;

\item $W^{DH}=\kappa W^{SP}$ is the ``leaking/dehydration field'', which is the spherical field $W^{SP}$ of the ion $q$ attenuated by a factor of $\kappa$;

\end{compactitem}

\noindent and we take the following equation for the dehydration energy (the key assumption for our model):
\begin{equation}
\Delta \mu^{DH}=%\Delta \mu^{SE}+
(\epsilon_p/\epsilon_w) \Delta \mu^{ZW}.
\end{equation}
Here, the shell-based dehydration energy $\Delta \mu^{ZW}$ is from a Laio/Zwolak-type semi-empirical approximation for the first shell energy $U_1^{DH}$ and for $\Delta \mu^{ZW}$ \cite{Zwolak:09}:
\begin{align}
U_1^{DH}&\approx \frac{q^2}{8\pi\epsilon_0\epsilon_p} %\left(\frac{1}{\epsilon_p}-\frac{1}{\epsilon_w}\right)
\frac{h_c}{R_{ion}(R_{ion}+h_c)}\\
%\Delta M_{0,X}^{DH}&=\Delta Q_{f,X}^{DH}=-(C_s/q)\Delta \mu_{X}^{DH} \\
\label{equ:shift_dh}
\Delta \mu_{X}^{ZW}&= U_1^{DH} \left(1- \sqrt{1-\left(R_c/R_{ion}\right)^2}\right)
\end{align}
where $X=\{K, Na,... \}$. This equation provides an expression for the hydration-shift in the selectivity of ion channels due to resonant quantized dehydration.

Fig.\ \ref{fig:DH1} illustrates the basis of the quantized  dehydration picture for monovalent ions in the  selectivity filters (SF) of NaChBac and relevant mutants. The radius of the SF is taken as $R_{c}=0.3$\,nm (the common value used in our simulations and calculations \cite{Kaufman:13a,Kaufman:13b}), whereas the first shell thickness $h_c=0.2$nm. Effective ionic radii $R_{}$ are taken from \cite{Shannon:76}. The dehydration energy $\Delta \mu_q^{DH}$ (blue solid line) is calculated to be proportional to $\sqrt{\delta R}$ where $\delta R= (R_{ion}+h_c) -R_c$, according to  \cite{Laio:99,Zwolak:09}. The shaded area indicates where ions retain their 1-st hydration shell within the SF. Plot (a) shows that Li$^+$ and Na$^+$ ions belong to ``fully hydrated'' range of $R_{ion},$ whereas K$^+$, Rb$^+$ and Cs$^+$ lie at a rapidly rising part of the dehydration energy curve and hence require significant shifts in $Q_f$ to balance dehydration penalty by site affinity to provide barrier-less conduction. %The Dashed arrows show putative shifts of resonant points for Na$^+$ and K$^+$ - selective channels.
Plot (b) shows the current $J_{ion}$ {\it vs.} $R_{ion}$ plotted for different values of $Q_f$, showing resonance for Na$^{+}$ and K$^{+}$, respectively. Monovalent ions present wide resonance curves, typical of weak ICB \cite{Kaufman:13b,Kaufman:15}.

Fig.\ \ref{fig:DH2} shows comparable plots for divalent ions.
Plot (a) shows that the Mg$^{2+}$ ion belong to the ``fully hydrated'' range of $R_{ion},$, whereas Ca$^{2+}$, Sr$^{2+}$ and Ba$^{2+}$ all require appropriate shifts in $Q_f$ to provide barrier-less conduction.
Plot (b) shows current $J_{ion}$ {\it vs} $R_{ion}$ curves for different values of $Q_f$, showing resonances for Mg$^{2+}$ and Ca$^{2+}$, respectively. The results and predictions are rather similar to those for monovalent ions but yield the much narrower resonances typical of strong ICB \cite{Kaufman:13b,Kaufman:15}, corresponding to the stronger dependence of conduction on ion radius.

In summary, after inclusion of resonant quantized dehydration effects, the resultant QD/ICB model predicts the following dependences of a channel's selectivity on $Q_f$, channel radius $R_c$ and ion size $R_{ion}$:

\begin{compactitem}

\item Narrow channels ($R_c\approx0.2$\,nm e.g.\ KscA channel) conduct fully dehydrated ions.  These channels tend to favour larger ions (K$^{+}$) (\cite{Laio:99,Corry:17}). When $Q_f$ is varied, narrow channels follow the original Eisenman rule, i.e.\ a highly charged SF tends to favour small ions \cite{Eisenman:83} and {\it vice-versa}.

\item Moderately-wide channels ($R_c\approx0.3-0.4$\,nm, e.g.\ NaChBac, NavAb, or Cav) conduct ions that retain their first hydration shells. In this case, low-charged mutants can conduct small (Li$^+$ and Na$^+$) ions and the growth of $|Q_f|$ leads to an {\it inverse} shift of Eisenman sequence toward the larger ions, i.e. Na$^+$ $\rightarrow$ K$^+$. We note, however, the earlier speculation by \citet{Corry:12} that the inverse shift might be due to a ``best fit" mechanism.

%\item Alternatively, the influence of $Q_f$ on K$^+$ vs Na$^+$ selectivity was putatively explained via a ``best fit" mechanism \cite{Corry:12}.

\end{compactitem}

\noindent We will now compare the predictions of the ICB/QD model with the experimental results. We will be especially interested in checking whether the theory can account for the Eisenman sequences obtained here and in earlier work \cite {Yue:02, Naylor:16, Guardiani:17}.

\subsection*{Ionic conductance for deleted-charge and balanced-charge mutants}
The ICB model predicts that, for an uncharged pore ($Q_f=0$), the self-energy barrier $U_q^{SE}$ should prevent conduction of any kind of ions. This condition corresponds to the ICB blockaded point $Z_0$ \cite{Kaufman:15}.

In the experiments, the mutations E191A  (generating LASWAS) and S192K (generating LEKWAS) were both used to produce NaChBac mutants with $Q_f$=0. These zero-charged mutants exhibited no measurable conduction of either monovalent nor divalent cations (see Supplemental Figure 2.), consistent with the ICB model. Although the model predicts even stronger current suppression for divalent ions, that difference was below the sensitivity of our measurements.

Rather similar ICB-driven blockade was recently observed in artificial sub-nm MoS$_2$ nanopores, where a voltage/energy gap was found corresponding to zero ionic current for both mono- and divalent ions \cite{Feng:16}.

%%------------------
%\begin{figure}[h]
%	\begin{center}
%		\includegraphics[width=\linewidth]{LASWAS.jpg}
%	\end{center}
%	\caption{ Original traces for LASWAS (A), LEKWAS (B) zero charge ($Q_f$=0) mutant channels recorded in the bath solution containing 140mM NaCl and LASWAS (C), LEKWAS (D) in 100mM CaCl$_2$ solution.}
%	\label {fig:laswas}
%\end{figure}
%%-----------------------------

%These results are consistent with ICB model. Model also predicts stronger suppression for divalent ions but that difference was below the sensitivity of our measurements.  %predicting that for $Q_f=0$ Coulomb blockade occurs (point $Z_0$) where high dielectric SE barrier prevents permeation of any ion through the channel.

\subsection*{Monovalent selectivity sequences of charged mutants}

Fig.\ \ref{fig:ed_mono} presents the mutation-induced transformations for the monovalent cation selectivity of the mutants studied. The peak conductivities for (a) wild type NaChBac LESWAS and the five mutants (b)-(f) as labelled were determined by normalising peak current magnitudes {\it from the same cell} recorded in a bath solution with Na$^+$, prior to replacement of the Na$^+$ by the test cation Li$^+$, K$^+$, Rb$^+$ or Cs$^+$ as indicated.

Fig.\ \ref{fig:ed_mono} can be thought as a ``mutation matrix'' in a Design of Experiment sense \cite{Kirk:82}. Taking Y=\{S, E, D\}, the different columns correspond to different residues at position 191: E191 (LEYWAS) for the left column and D191 (LDYWAS) for the right column. The different rows correspond to varying the residue in position 192 LXYWAS. The first row (a), (b) of Fig.\ \ref{fig:ed_mono} represents the singly-charged mutants LESWAS, LDSWAS, while the other two rows represent a 2$\times$2 submatrix of the nominally doubly-charged mutants LEEWAS/LDEWAS/LDEWAS/LDDWAS. Supplemental Material Fig.\ 3 shows the original $I-V$ characteristics for monovalent conductance. Supplemental Table 3 lists values for the permeability ($P_X/P_{Na}$ calculated from $E_{rev}$), and the conductance ($I_X/I_{Na}$ calculated from $I_{peak}$) ratios.

Fig.\ \ref{fig:ed_mono} shows the first observations of the expected ``shift phenomenon'': mutation E191D leads to a strong shift of the monovalent sequences from Na$^{+}$ (left column) towards K$^{+}$ (right column), whereas varying the residue in position 192 S$\rightarrow$E$\rightarrow$D leads to a weaker shift in the same direction. Now we consider these phenomena in more detail.

\subsection*{E$\rightleftharpoons$D substitution at key SF positions}

Earlier experiments have shown that bacterial channels/mutants exhibit different selectivity features depending on
similarly-charged residue rings (DDDD or EEEE) at the key site position 191 \cite{Yue:02,Naylor:16,DeCaen:14}
Studies of the E191D mutation in NaChBac \cite{DeCaen:14,Finol:14} (or, equivalent, E178D mutations in NavMs \cite{Naylor:16} and {\it vice versa}) have shown that the E$\rightleftharpoons$D substitution in the key position leads to a significant change in the selectivity features even though there is no change of the nominal charge.

Thus for NaChBac, comparison of the D191 mutants with the E191 wild-type channel shows that the residue at position 191 is the main determinant of monovalent selectivity. The mutation E191D leads to the emergence of K$^+$ conduction and to a general shift of monovalent selectivity sequences toward larger ions\cite{Finol:14}. A similar selectivity shift was recorded for the E178D mutation in the NavMs channel \cite{Naylor:16}.

Reversible  mutation-induced transformations of mono- and divalent selectivity sequences between Na$^+$-selective NaChaBac (LESWAS) and non-selective NsvBa (LDSWGS) bacterial channels was studied by \citet{DeCaen:14}. They showed that an E$\rightarrow$D mutation transforms the NaChBac channel to a non-selective analogue of NsVBa and {\it vice versa}, D$\rightarrow$E substitution at the SF of NsvBa results in an Na$^+$-selective channel similar to NaChBac, i.e.\ the selectivity is critically dependent on the type of charged residue at the key position inside the SF.

Our results (Fig.\ \ref{fig:ed_mono} (a), (b)) also demonstrate the strong influence of the E191D mutation on monovalent conduction and selectivity of the wild type LESWAS channel:

\begin{compactitem}

\item[(a)] LESWAS channel demonstrates features of sodium channel consistently with earlier observations \cite{Yue:02}

\item[(b)] The E191D LDSWAS conducts not only Li$^{+}$ and Na$^{+}$, but also K$^{+}$ ions, like the non-selective NaK channel, similar to the NsvBa (LDSWGS) channel \cite{DeCaen:14}.

\end{compactitem}

\noindent Comparison of the D191 mutants (left column of Fig.\ \ref{fig:ed_mono}) with appropriate E191 mutants (right column of Fig.\ \ref{fig:ed_mono}) shows that the residue in position 191 is the main determinant of monovalent selectivity in NaChBac mutants. Mutation E191D leads to shift of maximal monovalent conductance toward larger ions. It can be explained by significant growth of site affinity or ``equivalent charge" $Q_f^*$ connected with that substitution. Note that the NaK channel, possessing similar DDDD charged ring at the SF, is also Na/K non-selective \cite{Vora:08}.

Fig.\ \ref{fig:ed_mono} also shows that additional E192D mutations lead to relatively weak extra shift in the same direction.

\subsection*{Eisenman selectivity sequences: $Q_f$-induced shift}

\begin{figure}[t]
	\begin{center}
		\includegraphics[width=\gscale\linewidth]{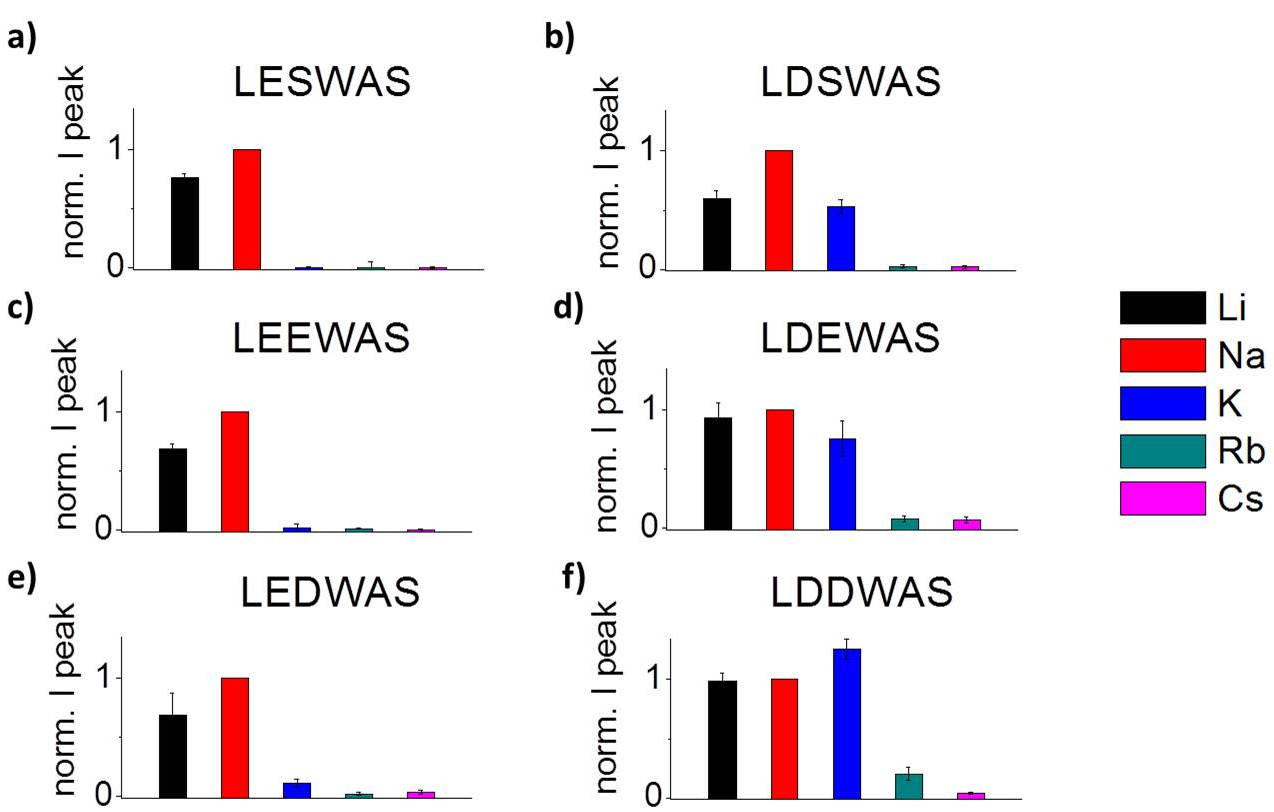}
	\end{center}
	\caption {(Color online). Monovalent cation permeability. Normalised peak currents $I_{\rm peak}$ for (a) wild type NaChBac LESWAS, (b) LDSWAS, (c) LEEWAS, (d) LDEWAS, (e) LEDWAS, and (f) LDDWAS mutant channels for Na$^+$, Li$^+$, K$^+$, Rb$^+$ and Cs$^+$ (as labelled) were determined by normalising peak current magnitudes recorded from the same cell in a Na$^+$ bath solution prior to replacement of extracellular Na$^+$ by the test cation. Averages ($\pm$SEM) are from at least 5 cells.}
	\label{fig:ed_mono}
\end{figure}

Fig.\ \ref{fig:ed_mono} presents Eisenman monovalent sequences for all the conducting mutants presented in the study. The ICB/QD model predicts a shift of the permeability ratio sequence towards the larger ions side, proportional to $|Q_f|$ value as main determinant of selectivity, Eq.\ \ref{equ:shift_dh}. However, the experiments reveal a more complicated picture. The mutant sequences can be separated into groups in terms of their dependence on $Q_f$ and on the residue type (D or E) at position 191, namely:

\begin{compactitem}

\item LESWAS, LEEWAS and LEDWAS. All mutants having an E191 residue present Na$^+$-centred Eisenman sequences, whereas increasing the nominal total charge from $|Q_f^{nm}|=4e$ for LESWAS to $|Q_f^{nm}|=8e$ for LEEWAS and LEDWAS leads only to a weak increase of K$^+$ permeability.

\item LDSWAS, LDEWAS and LDDWAS. The D191 mutants have Eisenman sequences shifted toward favouring K$^+$, and the difference in $Q_f$ values leads to weak additional permeability of large-sized ions in the LDEWAS and LDDWAS  mutants which have larger $|Q_f^{nm}|$.
	
\item LDDWAS. This double-ring ($|Q_f|=8e$) mutant exhibits an increased shift, in agreement with the ICB/QD model.

\end{compactitem}

%----------------------------------

\begin{figure}[t]
	\begin{center}
	\includegraphics[width=\gscale\linewidth]{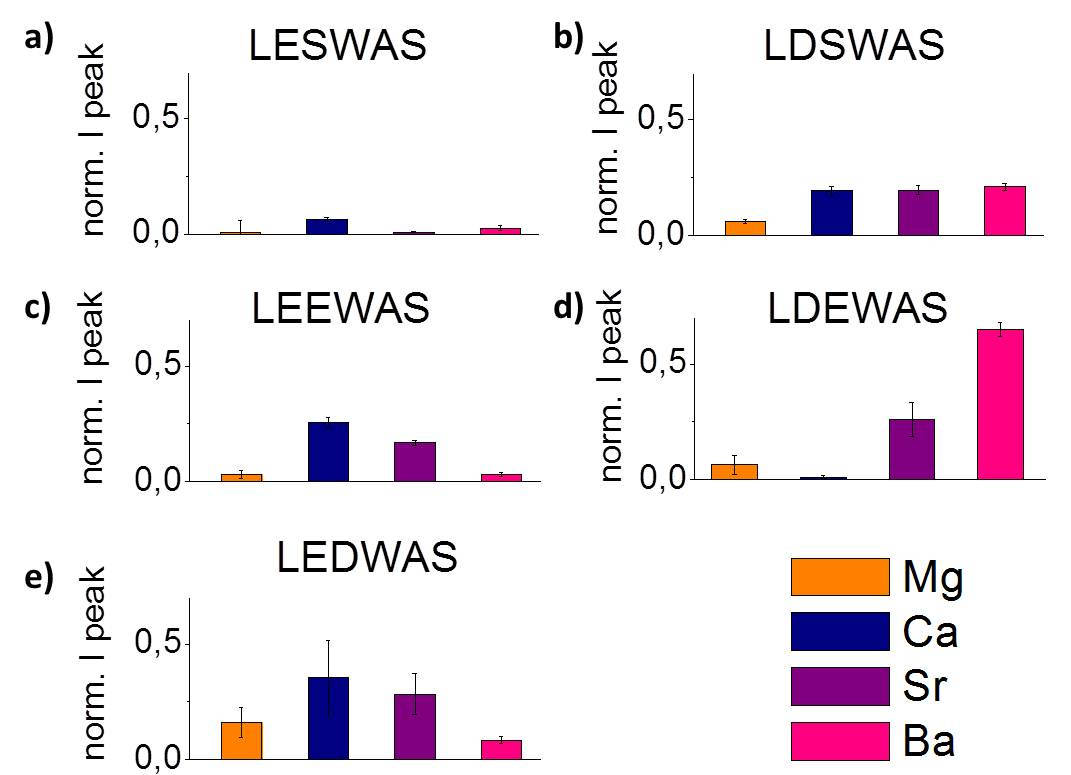}
	\end{center}
	\caption{(Color online.) Divalent cation permeability. Normalised peak currents $I_{\rm peak}$ for (a) wild type NaChBac LESWAS, (b) LDSWAS, (c) LEEWAS, (d) LDEWAS, and (e) LEDWAS mutant channels for Mg$^{2+}$, Ca$^{2+}$, Sr$^{2+}$ and Ba$^{2+}$ (as labelled) were determined by normalising peak current magnitudes recorded from the same cell in a Na$^+$ bath solution prior to replacement of extracellular Na$^+$ by the test cation. No divalent current could be measured for LDDWAS. Averages ($\pm$SEM) are from at least 5 cells. }
	\label{fig:ed_divalent}
\end{figure}

\subsection* {Divalent selectivity sequences of charged mutants}
Fig.\ \ref{fig:ed_divalent} presents the mutation-induced transformations of divalent cation permeability for the mutants studied. The peak conductivities for (a) wild type NaChBac LESWAS and for four of the mutant channels (b)-(e) are shown for the cations Mg$^{2+}$, Ca$^{2+}$, Sr$^{2+}$ and Ba$^{2+}$ (as labelled). As in the case of monovalent cations, they were normalised to the peak current recorded in Na$^+$ bath solution. No conduction of any divalent ion could be recorded for LDDWAS (Supplemental Fig.\ 4). Similarly to the monovalent case, Fig.\ \ref{fig:ed_divalent} can be thought as a ``mutation matrix" for divalent conduction. Columns represent the residue in position 191 (left for E191, right for D191) whereas rows show variation of the residue in position 192 (S192/E192/D192).

Supplemental Figure 4 shows the original $I-V$ characteristics for divalent conductance. Supplemental Table 3 lists values for the permeability ($P_X/P_{Na}$ calculated from $E_{rev}$), and for the conductance ($I_X/I_{Na}$ calculated from the $I_{peak}$) ratios.

Whereas the wild-type LESWAS channel exhibits little divalent cation permeability, in accordance with \cite{Yue:02} and \cite{Guardiani:17b}, most of the mutants show divalent conductivity.  Generally, the divalent selectivity sequences demonstrate a shift toward larger ions for both the E191D and S192D mutations. Thus LEEWAS and LEDWAS mutants show maximal permeability for Ca$^{2+}$  ions, whereas the LDEWAS mutant is Ba$^{2+}$-selective. Note that the E191D LDSWAS mutant conducts divalent ions, similarly to the NsvBa LDSWGS channel \cite{DeCaen:14}.

Remarkably, both the wild-type LESWAS channel ($Q_f=-4e$) and the doubly-charged ($Q_f=-8e$) LDDWAS mutant show strong blockade of all divalent ions, together with normal conductance for small monovalent ions (Na$^{+}$). We connect this phenomenon with ICB blockade points $Z_n$ of different-order: with smaller $n_1$ for LESWAS and larger $n_2>n_1$- for LDDWAS. The exact values of $n_1$ and $n_2$ will be discussed below.

	%-----------------------------------------------------------------------------
\begin{figure}[t]
	\center
	\includegraphics[width=0.8\linewidth]{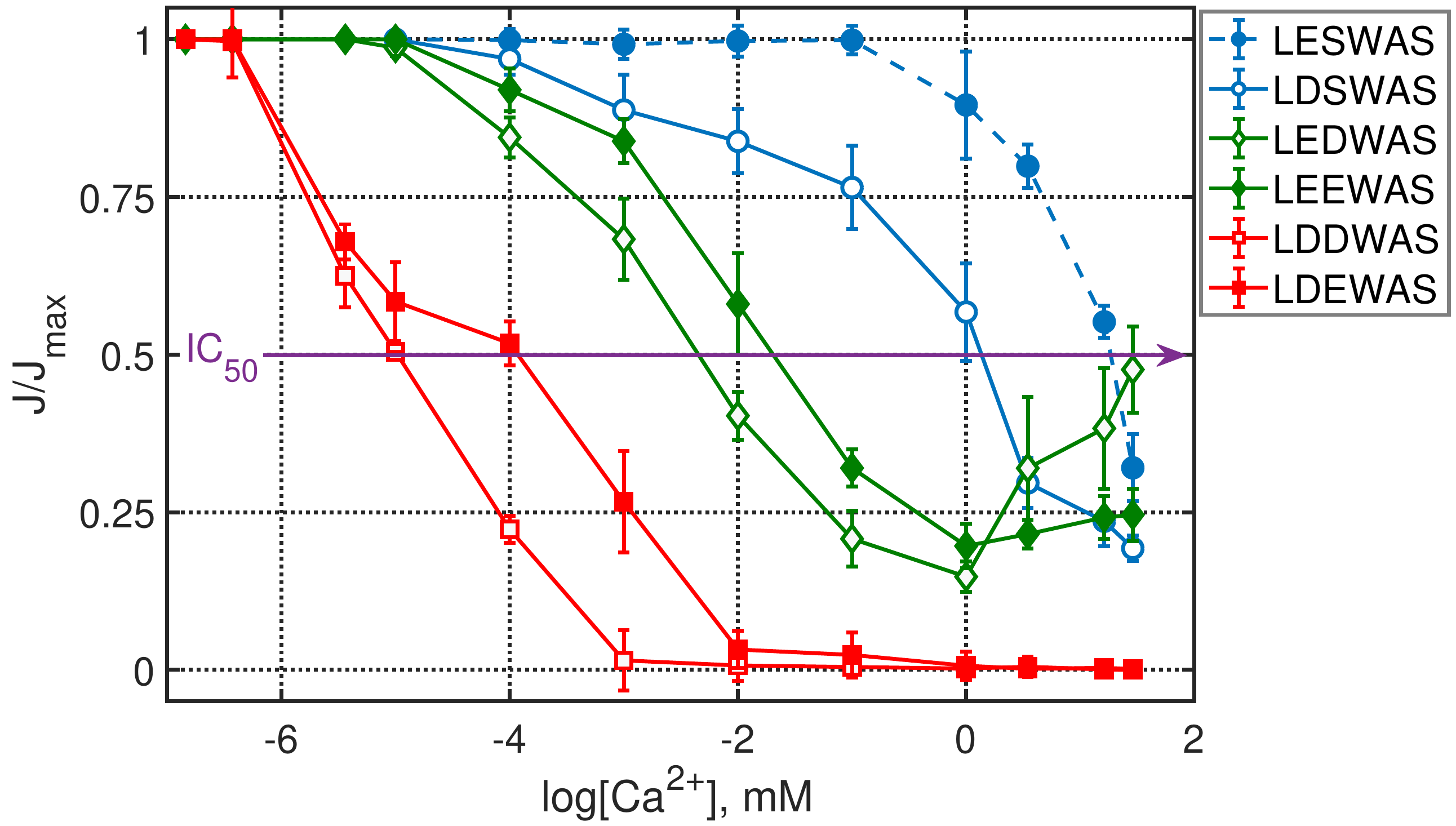}
	\caption{ %Demonstration of
		(Color online). Divalent blockade and  anomalous mole fraction effect (AMFE) in $Q_f^{nm}=-4e$ and $Q_f^{nm}=-8e$
		NaChBac mutants (indicated in inset). The relationships of the averaged normalized peak currents $J/J_{max}$ {\it vs.} $\log[Ca^{2+}]$ are plotted for increasing Ca$^{2+}$ content, ranging from 10 nM to 100 mM with Na$^+$ substitution. The line marked $IC_{50}$ shows the cross-section at $J/J_{max}=0.5$, i.e.\ Ca$^{2+}$ affinity.}
	\label{fig:dd_amfe}
\end{figure}
%------------------------------------------------------------------------------

\subsection* {Divalent blockade in charged mutants}
	
Strong divalent (e.g.\ Ca$^{2+}$) blockade of monovalent ( e.g. Na$^{+}$) currents followed by AMFE has been well-documented in calcium channels \cite {Sather:03,Nonner:98,Gillespie:08}. The study of AMFE provides a sensitive method of evaluating the divalent ions' affinity, which can be used to introduce an effective value of the fixed charge $Q_f^*$, as we will see below. Without the introduction of $Q_f^*$ the picture revealed by the scan of different cation permeabilities does not seem immediately explicable in terms of the extended ICB model using nominal values of $Q_f$. This study also helps us to understand the nature of the Ca$^{2+}$ interaction with the fixed charges in the pore and it is highly relevant to the model predictions.
	
Bath solutions containing mixtures of Na$^{+}$ and Ca$^{2+}$ in varying concentrations were used to investigate divalent blockade and the possibility of AMFE on both the whole cell current magnitude and current reversal voltages. The experiments started in a bath solution with 140 mM Na$^{+}$ and 10 nM of [Ca$^{2+}$]$_{free}$, which was sequentially replaced by solutions containing increasing addition of Ca$^{2+}$ at concentrations up to 1 mM [Ca$^{2+}$]$_{free}$, followed by solution in which Na$^{+}$ was replaced by Ca$^{2+}$ up to 28.7 mM of [Ca$^{2+}$]$_{free}$ ([Ca$^{2+}$]$_{total}$=100 mM).
Fig.\ \ref{fig:dd_amfe} shows the results of divalent blockade/AMFE experiments for NaChBac channels with the $Q_f^{nm}=-4e$ and $Q_f^{nm}=-8e$ mutants:

\begin{compactitem}

\item Notably, wild-type LESWAS channels do not exhibit any Ca$^{2+}$-dependent block of the Na$^+$ influx \cite{Kaufman:17a,Guardiani:17b}, and thus the reduced current results from extracellular Na$^+$ being replaced with equimolar Ca$^{2+}$ and represents the effect of substrate depletion.

\item LDSWAS shows moderately strong divalent blockade of Na$^+$ current with $IC_{50}\approx$1mM.

\item Na$^+$ currents through LEDWAS  \cite{Kaufman:17a} and LEEWAS were similarly sensitive to the presence of Ca$^{2+}$ and exhibited $IC_{50} \simeq 10\mu$M, with LEDWAS showing an additional weak shift relative to LEEWAS.

\item Ca$^{2+}$ blockade plots for LDEWAS and LDDWAS, were further  shifted toward lower concentrations relatively to the E191 mutants, with LDDWAS exhibiting a greater sensitivity to Ca$^{2+}$ than that for LDEWAS.

\end{compactitem}

\noindent The results show that the influence of the nominally equally-charged residues D and E on the Ca$^{2+}$ affinity (or on the $IC_{50}$ shift) depends systematically on both the residue type (D or E) and on the residue position (191 or 192). Note that the aspartate residue's contribution to the affinity is larger than the contribution of glutamate in the same position.

The difference between D and E also depends on position: it is significant for the E191D mutation (LESWAS {\it vs.} LDSWAS; LEDWAS {\it vs.} LDDWAS; LEEWAS {\it vs.} LDEWAS) and it is minor for the E192D mutation (LEEWAS {\it vs.} LDEWAS; LDEWAS {\it vs.} LDDWAS).

\subsection*{Effective charge $Q^*_f$ and Coulomb blockade oscillations}

We propose an {\it effective charge approach} to interpret the divalent blockade data in terms of the ICB picture. %It was shown in detail earlier \cite{Kaufman:15,Kaufman:17a}, that
The ICB model predicts quantitatively the Langmuir isotherm/Fermi-Dirac shape of the Na$^+$ current attenuation curve, and hence a linear dependence of the blockade threshold/affinity $\log IC_{50}$ on the fixed charge $Q_f$ \cite{Kaufman:15,Kaufman:17a}:
\begin{equation}
\log IC_{50} = a_Q+ b_Q   (Q_f^*/e),
\label {equ:q_eff}
\end{equation}
where the offset $a_Q$ and slope $b_Q$ are geometry-dependent coefficients, and $Q_f^*$ is now an effective value of $Q_f$.

Here, we take the inversion of Eq.\ \ref{equ:q_eff} as being the principal definition of $Q_f^*$. We also assume linearity: that any charged residue/ring has its own effective charge $Q_f^*$ , depending both on the type X,Y=\{D, E, S\} and position Pos=\{191, 192\} of the residue, so that Eq.\ \ref{equ:q_eff} is fulfilled for $Q_f^*$ in any Pos. The effective value of fixed charge $Q_f^*$ is an example of a general physical conception of ``effective values" (e.g.\ for diffusion coefficients or dielectric constants \cite {Kaufman:13c}) which allows us to use appropriate values outside their areas of formal applicability \cite{VanVechten:69}. Equation \ref{equ:q_eff} defines only the slope of $Q_f^*$ and not its absolute value. To establish the latter we use additional information coming from the Ca$^{2+}$ zero current positions, which we interpret as representing the $Q_f^*=Z_n\approx-2ne$ points of the stop bands; this interpretation is clearly non-unique and requires care.

%-------------------------------------
\begin{figure}[t]
	\begin{center}
		\includegraphics[width=\gscale\linewidth]{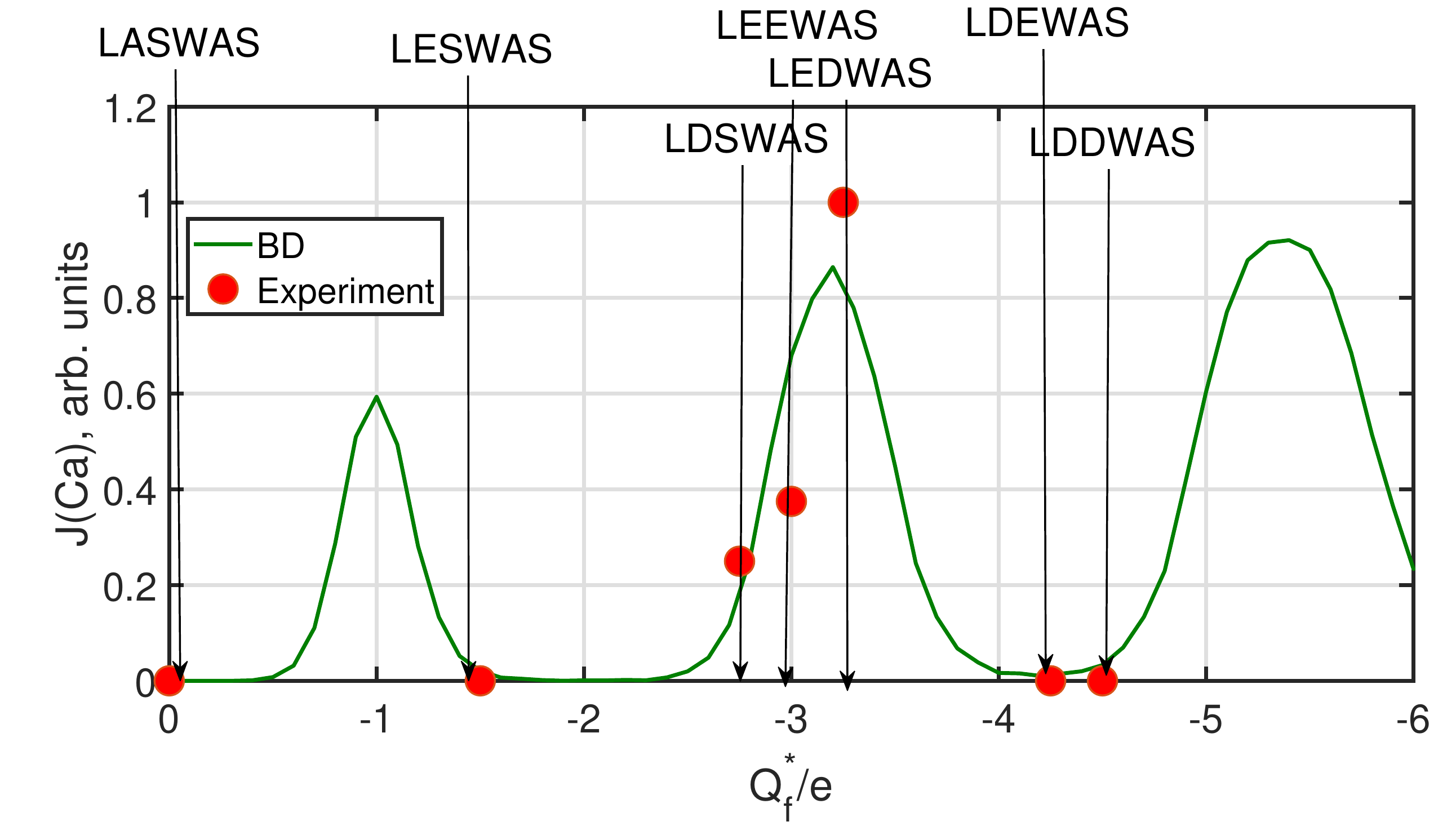}
	\end{center}
	\caption{(Color online). BD-simulated (green line) and experimental (red circles) multi-ion Ca$^{2+}$ conduction bands / ionic Coulomb blockade oscillations {\it vs} the effective fixed charge $Q_f^*$. The conduction bands in the Ca$^{2+}$ current $J$ were simulated for a pure bath with [Ca$^{2+}$]=160 mM. Experimental conductance data are taken from Fig.\ \ref{fig:ed_divalent}. The arrows show the $Q_f^*$ positions for the various NaChBac mutants calculated by fitting Eq.\ \ref{equ:q_eff} to AMFE data.}
	\label{fig:ca_bands_qeff}
\end{figure}
%------------------

Fig.\ \ref{fig:ca_bands_qeff} presents a putative $Q_f^*$ map for site-directed mutants combined with the multi-ion Ca$^{2+}$ conduction bands, as revealed by Brownian dynamics simulations and with appropriate experimental data superimposed. The vertical arrows indicate site-directed mutants with their respective effective fixed charges $Q_f^*$ deduced by the method described above.

The mutants' apparent charge mapping deviates significantly from the nominal $Q_f^{nm}$ values, but it allows us to explain the divalent conduction transformations that occur with growth of $Q_f^*$:

	\begin{compactitem}

\item LASWAS. $Q_f^*=0$. Zero conduction is predicted for all ions, due to $Q_f^*$ falling at the zeroth-order Coulomb blockade point $Z_0$, in agreement with the experiments.

\item LESWAS. $Q_f^*\approx -1.7e$. That is the $Z_1$ point for Ca$^{2+}$. Wild-type NaChBac is a sodium channel that should not conduct divalent ions. We assume that this must be because $Q_f^*$ is at a $Z_n\approx - z n e $ Coulomb blockade point. We take $n=1, Z_1\approx -2 e$. The alternative assumption ($n=2, Z_2\approx-4e$) does not agree with the experimental observation that $|Q_f^*({\rm D}191)|>|Q_f^*({\rm E}191)|$. The NaChBac sodium channel apparently possesses a significantly lower $|Q_f^*|<|Q_f^{nm}|$ than was thought previously \cite{Yue:02,Finol:14}. Interestingly, the bacterial sodium channel looks much closer to the mammalian Nav with its DEKA locus.

\item LDSWAS. $Q_f^*\approx -2.5e$. This should be a calcium-conducting non-selective mutant, which is in agreement with experiments. We use this point to fit $Q_f^*$(D191).

\item LEEWAS and LEDWAS, $Q_f^*\approx -3e$ and $-3.5e$ are calcium-conducting mutants. We infer that their $Q_f^*$ values must lie close to the $M_1$ resonant point for Ca$^{2+}$. The model used for the BD simulations takes no account of quantized dehydration effects, which is probably why the agreement with experiment is not very good.

\item LDEWAS and LDDWAS, $Q_f^*\approx -4e$ and $-4.5e$, $Z_2$ Ca$^{2+}$ point. Experimentally, these mutants do not conduct Ca$^{2+}$ ions (LDDWAS does not conduct divalent ions at all), so we assume them to be close to the nearest Ca$^{2+}$ Coulomb blockade point, i.e.\ $Z_2$. This point was then used to scale the $Q_f^*$ map.

\end{compactitem}

\noindent Both the BD simulations and experimental Ca$^{2+}$ conduction points show blockaded zero-current bands $Z_n$ separated by conduction bands $M_n$, in agreement with the basic ICB model \cite {Kaufman:15}.

Figure \ref{fig:ca_bands_qeff} represents the first observation of ICB oscillations of Ca$^{2+}$ conduction in biological ion channels. The oscillations manifest themselves strongly at the room temperature, unlike their electronic counterpart in quantum dots which become significant only at low temperatures \cite{Averin:86}. We comment that $Q_f^*$ constitutes the main determinant of selectivity, in agreement with the ICB/QD model.

%-----------------------------------------------------------------------------
\begin{figure}[t]
	\center
	\includegraphics[width=0.9\linewidth]{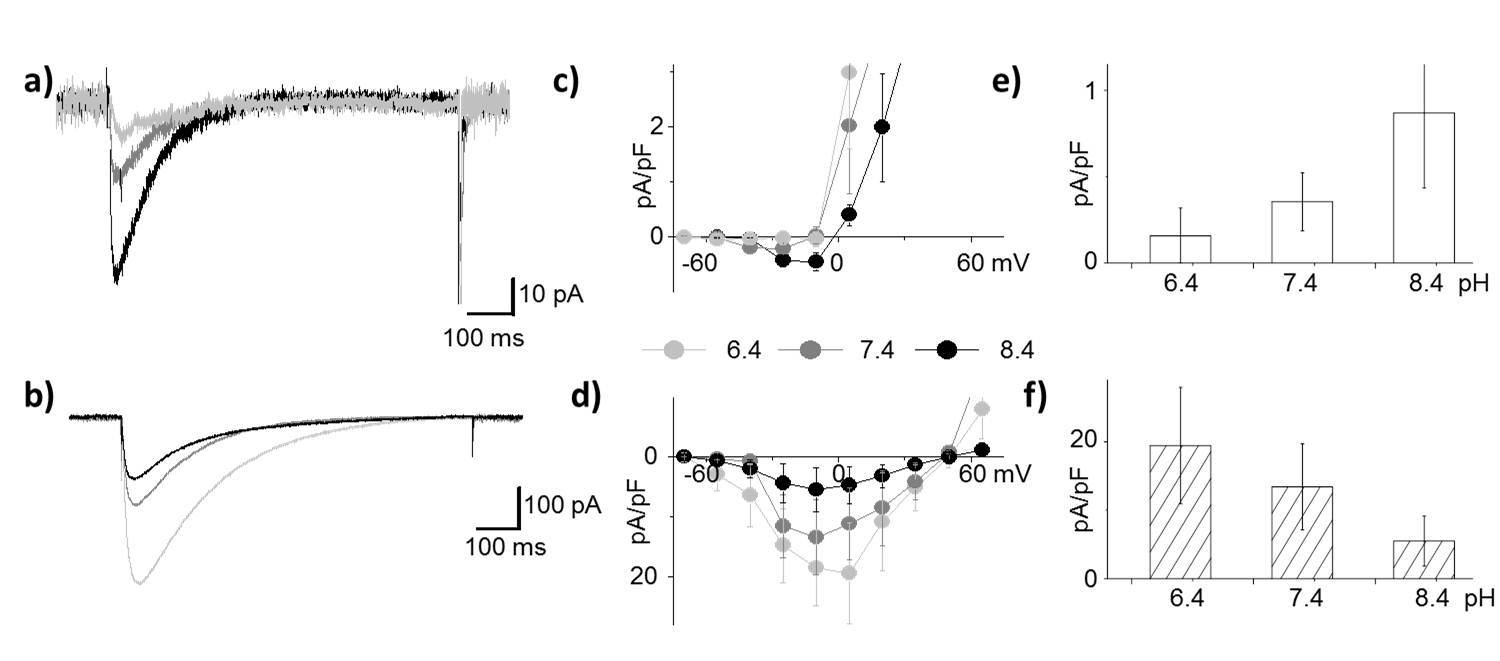}
	\caption{(Color online). Effect of extracellular pH on Ca$^{2+}$ currents from LESWAS (top panels) and LDSWAS (bottom panels) NaChBac channels. Original whole cell current recordings from cells expressing (a) LESWAS  and (b) LDSWAS in response to a depolarising step to -10 mV (from $V_{\rm hold} = -100\,$mV) in SBS containing 100\,mM Ca$^{2+}$  at pH 6.4 (light grey), 7.4 (dark grey) and 8.4 (black). (c, d) Mean (+/- SEM) current-voltage relationships of whole-cell currents as described for parts (a) and (b). (e, f) Peak current densities obtained from current-voltage relationships shown in parts (c) and (d).
	}
	\label{fig:ph_mutants}
\end{figure}
%------------------------------------------------------------------------------

\subsection*{Site protonation model of effective charge $Q_f^*$}
\label{sec:proton}
%-----------------------------------------------------------------------------
We now propose that the difference between the nominal $Q_f^{nm}$ and effective $Q_f^*$ values of $Q_f$ is connected to different protonation of the charged residues inside DDDD and EEEE rings \cite{Furini:14,Corry:12}.

The negative fixed charge $Q_f$ of bacterial channels/mutants is provided by the ionised side chains of the aspartate (D) and glutamate (E) residues, which  are the only negatively-charged protein side chains. They have the same nominal charges ($Q_f^{nm}$=-1) and very similar pK$\approx$4 but different lengths, providing the EEEE-ring with a smaller radius $R_Q^{E}$ than $R_Q^{D}$ of the DDDD-ring. This difference may lead to different (and putatively opposite) effects: arising from the difference in local binding, and from a difference in protonation.

We point out that there is also a possible alternative explanation to protonation. Recent MD studies \cite{Fedorenko:18b} suggest that, when aspartate is in position 191, its charged side chains point {\it away from} the centre of the channel. Consequently, an incoming Ca$^{2+}$ ion might ``feel'' an effective charge that is lower than the nominal one.

We hypothesize, however, that protonation is the dominant effect and that the increase of affinity related to E191D mutation is connected to significant protonation of the relatively small-radius EEEE charged ring due to overlapping of the electron and proton clouds between neighbouring residues in the ring: such effects have been studied by \citet{Furini:14}.
Similar effects due to space restrictions on the pKa and protonation state of glutamate were also calculated for the narrower KcsA channel, where pKa was shifted to pKa=9.2 \cite{Berneche:02}.

Otherwise, the larger DDDD ring could be more ionized at physiological pH, closer to  the full ionization of free residues.
The protonation-related interpretation implies that the effective $Q_f^*$ value is not averaged "conceptual value" but it reflects the ``true'' electrostatic value of $Q_f$, differing from the nominal $Q_f^{nm}$ value corresponding to the ``ideal'' maximum possible value of $Q_f$.

It had been shown previously that the pH in the external solution can alter dramatically the Na$^+$ conductance for NaChBac \cite{DeCaen:14} and Na$^+$/K$^+$ selectivity \cite{Finol:14}; there are, however, no data about the dependence of the Ca$^{2+}$ selectivity on pH.

Fig.\ \ref{fig:ph_mutants} presents the results of our preliminary study of the effect of variations in extracellular pH on Ca$^{2+}$ conductance in the LESWAS and LDSWAS NaChBac channels. These effects clearly depend on residue type at position 191. Panel (e) shows that the wild type LESWAS-mediated inward Ca$^{2+}$ current was small and relatively insensitive to pH changes. Such behaviour corresponds to the position of LESWAS on the ICB conduction {\it vs.}\ $Q_f^*$ map (Fig.\ \ref{fig:ca_bands_qeff}), i.e.\ to the Ca$^{2+}$ stop band for all pH, thus confirming our interpretation of LESWAS as lying at the Ca$^{2+}$ blockade point. Panel (f) for the LDSWAS mutant demonstrates a significant calcium current $I$, decreasing with growth of pH (and $Q_f^*$) , which corresponds to the decreasing-slope side of the ICB oscillation $I$ {\it vs} $Q_f^*$. Such behaviour is inconsistent with our mapping (Fig.\ \ref{fig:ca_bands_qeff}) in which the LDSWAS ``working point'' is located on the increasing-slope side. This contradiction requires further investigation.

\section*{Conclusions}
\label{sec:concl}
Our application of site-directed mutagenesis and whole-cell patch-clamp technique to investigate the influence of the fixed charge at the SF of the NaChBac bacterial channel has enabled us to explain several hitherto puzzling features of valence and alike-charge selectivity. It was shown that the experimental results can be accounted for within the framework of the novel ICB/QD model, and that the latter is able to describe alike-charge selectivity as well as valence selectivity.

In particular, by systematically changing residues D and E in the key positions 191, 192, and measuring the conduction and selectivity of the resultant LASWAS, LEKWAS, LESWAS, LDSWAS, LEEWAS, LEDWAS, LDEWAS and LDDWAS mutants:

\begin{compactitem}

\item[1.] The systematic measurements of the Na$^+$/Ca$^{2+}$ divalent blockade threshold, and hence of Ca$^{2+}$ affinity, necessitated the introduction of an effective value $Q_f^*$ of the fixed charge in order to be able describe the results.  It was found to be significantly smaller than the nominal fixed charge $Q_f$. It enables all of the experimental results to be understood and it can be considered as the main determinant of selectivity for bacterial channels.

\item[2.] Our new ICB/QD model, applied to the results of mutation studies on the NaChBac channel, predicts that increase of $Q_f^*$ should lead to a shift in the Eisenman selectivity sequences toward ions with larger radii, for both monovalent and divalent cations. The experiments revealed that such shifts occur in reality.

\item[3.] The charge-varied mutants exhibit divalent blockade and anomalous mole fraction effect in solutions containing mixtures of Na$^{+}$ and  Ca$^{2+}$. However, mutants with the same nominal $ Q_f$ (-4 or -8), generated with different combinations of glutamate and aspartate at the SF, exhibit markedly different cation permeation profiles.

\item[4.] The zero-current Ca$^{2+}$ Coulomb-blockaded point $Z_2\approx-4e$ was observed directly for the LDEWAS mutant., together with $Z_1\approx-2e$ point for LESWAS and resonant conduction point $M1\approx-3e$ for LEDWAS. This result amounts to the first experimental observation of ICB oscillations in biological ion channels, {\it cf.} the recent report of its observation in artificial sub-nm nanopores \cite{Feng:16}.

\item[5.] The $Q_f^*$ values for aspartate and glutamate rings are found to be different, and they also differ according to the ring's position along the SF. We putatively connect this difference with difference ionisation/protonation between the EEEE and DDDD charged rings, which should also depend on their radii. Another possibility \cite{Fedorenko:18b} is that the EEEE and DDDD rings are in different conformational states.
    
\item[6.] Our preliminary study of the effects of extracellular pH in the 6.4-8.4 range on Ca$^{2+}$ conductance confirms our attribution of the Ca$^{2+}$ $Z_1$ ICB blockade point for LESWAS, whereas the LDWAS behavior does not fit the scheme and clearly requires farther investigation.

\end{compactitem}
\noindent The overall conclusion is that the ICB/QD model, allowing for possible protonation, provides a good  description of most features of the conduction and selectivity of the NaChBac channel and its mutants. It is reasonable to suppose that the model will be equally applicable to the many other ion channels that are governed by the same general physical principles.

%}

%\section*{Authors Contribution}
%\begin{compactitem}
%\item[{\bf OAF:}]
%Experimental design, conducting experiments, text writing/editing, data analysis.
%\item [{\bf IKK:}]
%Experimental design, model development, text writing/editing, data analysis.
%\item [{\bf WATG:}]
%Model development, data analysis
%\item [{\bf DGL:}]
%Model development, data analysis
%\item[{\bf CG}]
%MD simulations, data analysis
%\item[{\bf IKh}]
%MD simulations, data analysis
%\item[{\bf SKR:}]
%Experimental design, data analysis, text writing/editing
%\item[{\bf PVEMcC:}]
%Model development, data analysis, text writing/editing
%\end{compactitem}

\section* {Acknowledgements}

The research was supported by the Engineering and Physical Sciences Research Council UK ( grant No. EP/M015831/1 ) and by a Leverhulme Trust Research Project Grant RPG-2017-134.
We are grateful to
M.\ L.\ Barabash,
R.\ S.\ Eisenberg,
C.\ Guardiani,
A.\ Stefanovska, and
M.\ Di Ventra,
for comments and useful discussions.
We thank Huaping Sun, for her much appreciated help with the generation of mutant channels.

%%\thebibliography{
\renewcommand{\bibnumfmt}[1]{{#1}.}
\def\bibfont{\small}
%%}

\bibliographystyle{biophysj}
%\bibliographystyle{unsrtnat}

%\bibliographystyle{abbrvnat}
%\bibliography{ionchannels}

%\bibliography{i}



\section*{Supplementary material}

# Quantized Dehydration and the Determinants of Selectivity in the NaChBac Bacterial Sodium Channel

O. A. Fedorenko, I. Kh. Kaufman, W. A. T. Gibby, D. G. Luchinsky, S. K. Roberts,

P. V. E. McClintock

## Supplemental information.

**Supplemental table 1.** Oligonucleotides containing the sequence for the desired amino acid substitutions used in this study.

| Amino acid sequence | Nominal Q$f$ | Template | Forward primer | Reversed primer |
|---|---|---|---|---|
| L**A**SWAS | 0 | LESWAS | GGTCACGCTAgccTCATGGGCGAGcggcg | ACTTGGAACAATGTTAACAAACtaagc |
| LE**K**WAS | 0 | LESWAS | CACGCTAGAGaagTGGGCGAGCG | ACCACTTGGAACAATGTTAAC |
| L**D**SWAS | -4 | LESWAS | GGTCACGCTAgatTCATGGGCGAG | ACTTGGAACAATGTTAACAAACTAAGC |
| LE**E**WAS | -8 | LESWAS | CACGCTAGAGgagTGGGCGAGCG | ACCACTTGGAACAATGTTAAC |
| LE**D**WAS | -8 | LESWAS | CACGCTAGAGgatTGGGCGAGCG | ACCACTTGGAACAATGTTAAC |
| L**DE**WAS | -8 | LEEWAS | GGTCACGCTAgatGAGTGGGCGA | ACTTGGAACAATGTTAACAAACTAAGCTG |
| L**DD**WAS | -8 | LEDWAS | GGTCACGCTAgatGATTGGGCGA | ACTTGGAACAATGTTAACAAACTAAGC |

## Electrophysiology

Whole-cell voltage clamp recordings were performed at room temperature (20 °C) using an Axopatch 200A (Molecular Devices, Inc.) amplifier. Whole cell currents were elicited by a series of step depolarizations (+95mV to -85mV in -15mV steps) from $V_{hold}$ of -100mV.

Patch-clamp pipettes were pulled from borosilicate glass (Kimax, Kimble Company, USA) to resistances between 2-3 MOhm. Pipette solution was either PS1 (120mM Cs-methanesulfonate, 20mM Na-gluconate, 5mM CsCl, 10mM EGTA, and 20mM HEPES, pH7.4 adjusted with 1.8mM CsOH) or Cs-free PS2 (15mM Na-gluconate, 5mM NaCl, 90mM NMDG, 10mM EGTA, and 20mM HEPES, pH7.4 adjusted with 3mM HCl). Unless otherwise stated, GOhm seals were obtained in standard bath solution ($SBS_{Na}$; 140mM Na-methanesulfonate, 5mM CsCl, 10mM HEPES and 10mM glucose, pH=7.4 adjusted with 4.8mM CsOH). For experiments with LDDWAS and LDEWAS, Cs-free $SBS_{Na}$ ($SBS_{Nax}$: 132mM Na-methanesulfonate, 5mM NaCl, 10mM HEPES and 10mM glucose, pH=7.4 adjusted with 3.6mM NaOH) was used; these channels are permeable to Cs and thus its removal from the test solutions was required to establish pseudo-bionic conditions (Supplemental Fig.1). Furthermore, the bath solutions ($SBS_{Nax}$ and $SBS_{Nax}$ in which 140 mM $Na^+$ was replaced with the test monovalent cation) used for recording LDDWAS activity in the presence of extracellular monovalent cations also contained 10 mM EGTA (see below) to account for the high affinity Ca2+ blockade of the channel (Figure 5).

Permeability to different test cations was determined by replacing 140mM NaCl with equimolar test monovalent cation or 100mM test divalent cation using Cl- salts. Permeability ratios ($P_x/P_{Na}$) were determined from whole cell current reversal potentials ($E_{rev}$) for monovalent/divalent cations according to [34, 35]. The effective activity coefficients were calculated using the Debye-Hückel equation (Supplemental Table 2).

In experiments with varied Na/Ca mixtures, GOhm seals were first obtained in $SBS_{Na}$ (or $SBS_{Nax}$ for experiments with LDDWAS and LDEWAS) containing 10 nM $[Ca^{2+}]_{free}$, and the bath solution replaced with solutions containing increasing concentrations of $Ca^{2+}$. EGTA was used to fix $Ca^{2+}$ concentrations at 10 nM, 100 nM *and 1* μM, and HEDTA to fix Ca2+ concentrations at 10 μM, 100 μM and 1 mM. $[Ca^{2+}]_{free}$ were calculated using Webmaxc (ttp://web.stanford.edu/cpatton/webmaxcs.htm). $Na^+$ and $Ca^{2+}$ (for 10mM, 50mM and 100mM) activities were calculated using the Debye-Hickel equation (we consider $[Ca^{2+}]_{free}$ = ion activity of $Ca^{2+}$); for details see Supplemental Table 2.

Osmolarities of all solutions were measured using a Wescor vapor pressure osmometer (model 5520) and adjusted to 280 lmOsmKg-1 using sorbitol. All solutions were filtered with a 0.22mm filter before use. Whole cell currents were recorded 3 minutes after obtaining whole cell configuration to ensure complete equili- bration of the pipette solution and cytosol. The bath solution was grounded using a 3 M KCl agar bridge; the liquid junction potential, determined experimentally [36], agreed with that calculated (using JPCalc program, Clampex, Axon Instruments, Inc.), were less than 2.6mV, and were not accounted for.

The recording chamber volume was approximately 200 μl and was continuously exchanged by a gravity-driven flow/suction arrangement at rate of ≈2ml/min; to ensure complete exchange of bath solution. Electrophysiological recordings were initiated after >4 minutes of continuous solution change.

Results were analysed using Clampfit 10.1 software (Molecular Devices) and Origin9.1 (OriginLab). Data are presented as means ± SEM (n), where n is the number of independent experiments.

**Supplemental table 2.** Concentrations and the effective activity coefficients of solutions used for permeability ratios determination.

|  | $Na^+$ | $Li^+$ | $K^+$ | $Rb^+$ | $Cs^+$ | $Mg^{2+}$ | $Ca^{2+}$ | $Sr^{2+}$ | $Ba^{2+}$ |
|---|---|---|---|---|---|---|---|---|---|
| Concentration (in mM) | **140** | **140** | **140** | **140** | **140** | **100** | **100** | **100** | **100** |
| activity coefficients (Debye-Hückel equation) | 0.74 | 0.78 | 0.72 | 0.71 | 0.71 | 0.34 | 0.29 | 0.25 | 0.25 |
| Free concentrations/ activities | 103.6 | 109.2 | 100.8 | 99.4 | 99.4 | 34 | 29 | 25 | 25 |

**Supplemental table 3.** Total and free concentrations of main cations and chelators in solutions used in the study.

| Total concentrations | | | Free concentrations/ ion activities | |
|---|---|---|---|---|
| Na$^+$ | Ca$^{2+}$ | Chelator | Na$^+$ | Ca$^{2+}$ |
| 140 | 0.23 mM | 2 mM EGTA | 110.6 mM | 10 nM |
| 140 | 1.14 mM | 2 mM EGTA | 110.6 mM | 100 nM |
| 140 | 1.86 mM | 2 mM EGTA | 110.6 mM | 1 µM |
| 140 | 0.69 mM | 1 mM HEDTA | 110.6 mM | 10 µM |
| 140 | 1.05 mM | 1 mM HEDTA | 110.6 mM | 100 µM |
| 138 | 2 mM | 1 mM HEDTA | 108.7 mM | 1 mM |
| 130 | 10 mM | 1 mM HEDTA | 101.0 mM | 4.13 mM |
| 70 | 50 mM | 1 mM HEDTA | 52.2 mM | 18.3 mM |
| 40 | 100 mM | 1 mM HEDTA | 28.4 mM | 31.7 mM |

**Supplementary table 4.** Reversal potentials ($E_{rev}$), permeability ratios ($P_X/P_{Na}$) and the relative peak inward current ($I_X/I_{Na}$) for the tested cations for the channels with $Q_f = -4$ and $Q_f = -8$. All values are means (± SEM) with the number of experiments in parenthesis. $E_{rev}$ is the mean reversal potential in millivolts measured for each cation; in cases where inward current was not detected, estimated values for $E_{rev}$ were determined as voltage at which outward current could be detected. Permeability ratios ($P_X/P_{Na}$) for each cation were calculated (see in Methods) from paired changes in $E_{rev}$ measured for a given cell perfused first with control $Na^+$ solution and after replacement with the test cation solution according to the following equations (*Hille, 1972*; *Sun et al., 1997*) for monovalent cations:

$P_X/P_{Na} = \alpha Na_i / \alpha X_e [\exp(\Delta E_{rev}/(RT/F))]$,

and divalent cations:

$P_Y/P_{Na} = \{\alpha Na_i [\exp(E_{rev} F/RT)][\exp(E_{rev} F/RT)+1]\}/4\alpha Y_e$,

where $\Delta E_{rev}$ is the change in reversal potential on replacing $Na^+$ with the tested cation, α is the activity coefficients for the ion (i, internal and e, external), R - the universal gas constant, T - absolute temperature, and F - the Faraday constant. The effective activity coefficients (αx) were calculated using the Debye-Hückel equation and are listed in Supplemental Table 2. $I_X/I_{Na}$ was measured as the ratio of maximum peak inward current observed for the test cation to that for observed $Na^+$ in the same cell.

| | | LESWAS | LDSWAS | LEEWAS | LDEWAS | LEDWAS | LDDWAS |
|---|---|---|---|---|---|---|---|
| **$Li^+$** | $E_{rev}$ | 48.7±1.42 mV (n=6) | 42.3±1.8 mV (n=6) | 47.9±0.4 mV (n=9) | 43.3±2.1 mV (n=6) | 48.2±1.57mV (n=6) | 42.3±0.7 mV (n=7) |
| | $P_{Li}/P_{Na}$ | 0.8 | 0.6 | 0.7 | 0.6 | 0.8 | 0.6 |
| | $I/I_{Na}$ | 0.8 | 0.6 | 0.7 | 0.9 | 0.7 | 1 |
| **$Na^+$** | $E_{rev}$ | 51.0±1.26 mV (n=17) | 49.0±1.0 mV (n=21) | 51.8±1.1 mV (n=16) | 47.8±1.4 mV (n=10) | 50.2±0.81mV (n=14) | 46.4±1.0 mV (n=14) |
| | $P_{Na}/P_{Na}$ | 1 | 1 | 1 | 1 | 1 | 0.9 |
| | $I/I_{Na}$ | 1 | 1 | 1 | 1 | 1 | 1 |
| **$K^+$** | $E_{rev}$ | <-30 mV (n=8) | 42.5±2.4 mV (n=5) | -15.9±3.8 mV (n=7) | 43.4±3.1 mV (n=7) | 9.7±2.31 mV (n=8) | 48.5±0.6 mV (n=7) |
| | $P_K/P_{Na}$ | <0.1 | 0.8 | <0.1 | 0.8 | 0.2 | 1 |

|  |  |  |  |  |  |  |  |
|---|---|---|---|---|---|---|---|
|  | I/I$_{Na}$ | <0.1 | 0.6 | <0.1 | 0.8 | 0.1 | 1.2 |
| **Rb$^+$** | E$_{rev}$ | <-30 mV (n=5) | -8.8±6.0 mV (n=5) | -22.9±4.7 mV (n=6) | 10.3±1.5 mV (n=5) | -7.9±1.57 mV (n=6) | 35.9±4.1 mV (n=7) |
|  | P$_{Rb}$/P$_{Na}$ | <0.1 | 0.1 | <0.1 | 0.2 | 0.1 | 0.6 |
|  | I/I$_{Na}$ | <0.1 | <0.1 | <0.1 | 0.1 | <0.1 | 0.2 |
| **Cs$^+$** | E$_{rev}$ | <-30 mV (n=5) | -29.0±5.0 mV (n=5) | -40.4±6.7 mV (n=6) | -3.9±3.1 mV (n=5) | -11.8±7.42 mV (n=6) | 8.2±1.1 mV (n=7) |
|  | P$_{Cs}$/P$_{Na}$ | <0.1 | <0.1 | <0.1 | 0.1 | 0.1 | 0.2 |
|  | I/I$_{Na}$ | <0.1 | <0.1 | <0.1 | 0.1 | <0.1 | <0.1 |
| **Mg$^{2+}$** | E$_{rev}$ | -26.3±9.24mV (n=5) | 19.0±3.0 mV (n=6) | -3.9±2.2 mV (n=6) | 15.9±7.4 mV (n=5) | 49.5±1.78 mV (n=6) | Block (n=15) |
|  | P$_{Mg}$/P$_{Na}$ | 0.1 | 0.7 | 0.2 | 0.6 | 6.5 | - |
|  | I/I$_{Na}$ | <0.1 | <0.1 | <0.1 | 0.1 | 0.2 | - |
| **Ca$^{2+}$** | E$_{rev}$ | -5.1±2.63 mV (n=9) | 53.3±1.2 mV (n=7) | 58.7±1.8 mV (n=9) | block | 71.5±1.26 mV (n=11) | Block (n=16) |
|  | P$_{Ca}$/P$_{Na}$ | 0.2 | 10.2 | 14.1 | - | 41.1 | - |
|  | I/I$_{Na}$ | <0.1 | 0.4 | 0.25 | - | 0.4 | - |
| **Sr$^{2+}$** | E$_{rev}$ | <-30 mV (n=5) | 54.8±2.1 mV (n=6) | 45.0±3.8 mV (n=7) | 75.8±1.8 mV (n=7) | 59.2±4.34 mV (n=7) | Block (n=15) |
|  | P$_{Sr}$/P$_{Na}$ | <0.1 | 13.2 | 8.5 | 66.8 | 18.5 | - |
|  | I/I$_{Na}$ | <0.1 | 0.4 | 0.2 | 0.3 | 0.3 | - |
| **Ba$^{2+}$** | E$_{rev}$ | <-30 mV (n=5) | 40.9±3.8 mV (n=6) | 13.4±0.8 mV (n=8) | 61.6±2.5 mV (n=5) | 49.0±1.98 mV (n=9) | Block (n=15) |
|  | P$_{Ba}$/P$_{Na}$ | <0.1 | 4.7 | 0.7 | 22.2 | 8.5 | - |
|  | I/I$_{Na}$ | <0.1 | 0.2 | <0.1 | 0.1 | 0.1 | - |

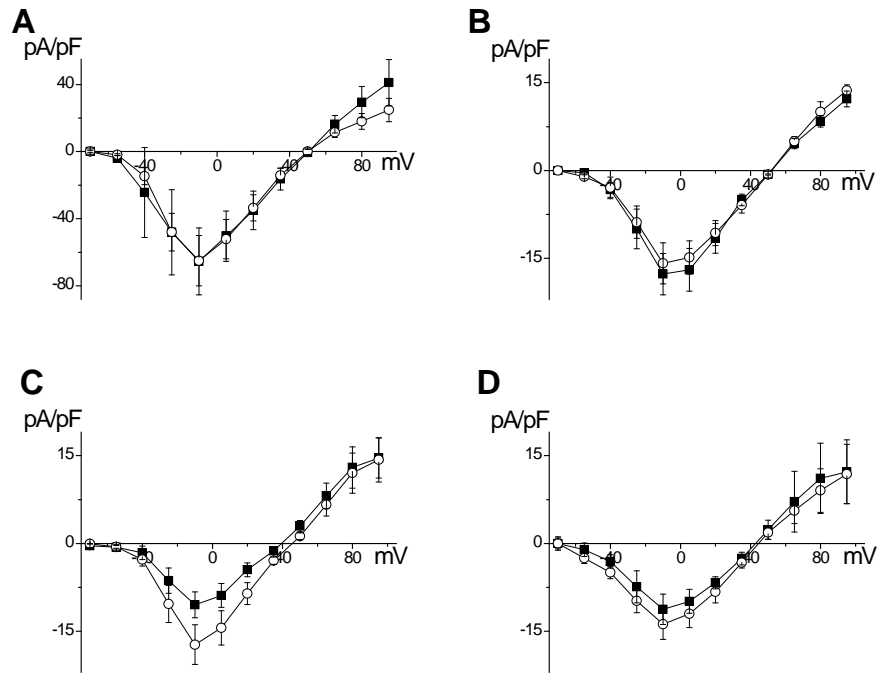

**Supplemental Figure 1**. Optimising the pipette solution. Plot of average peak current ($I_{peak}$) density against test voltage from cells expressing LDSWAS (A; ■ - n =10; ○ - n = 15; note the scale); LEEWAS (B; ■ - n =16; ○ - n = 7), LDDWAS (C; ■ - n =7; ○ - n = 14) and LDEWAS (D; ■ - n =10; ○ - n = 10) mutant channels recorded in SBS_$Na_x$ and PS1 (■) or PS2 (○) in response to test voltages ranging from +95 mV to -70 mV (in -15 mV steps) from $V_{hold}$ -100 mV. Note that for LDDWAS (and, to a lesser extent, LDEWAS) the removal of $Cs^+$ (○) from the pipette solution resulted in a shift in $E_{rev}$ and larger inward current consistent with $Cs^+$ permeation.

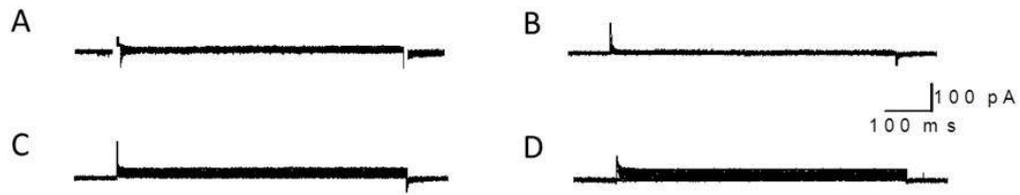

**Supplemental Figure 2.** Original traces for LASWAS (A, B) and LEKWAS (C, D) mutant channels recorded in the bath solution (SBS$_{Na}$)containing 140 mM NaCl (A, C) or in 100 mM CaCl$_2$ solution (B, D) and PS1 in response to test voltages ranging from +95 mV to -70 mV (in -15 mV steps) from V$_{hold}$ -100 mV.

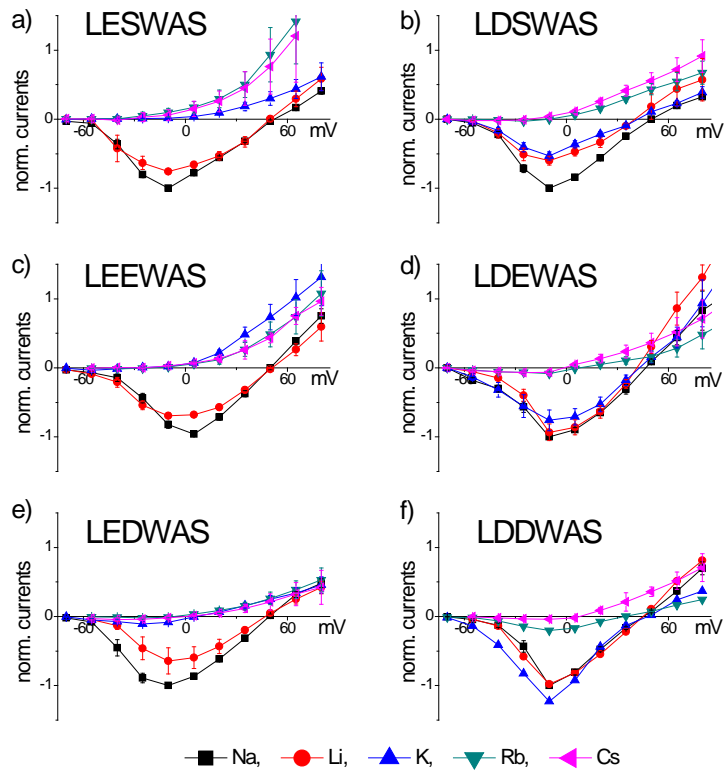

**Supplemental Figure 3.** Monovalent cation permeability. Mean peak current voltage relationships for wild type NaChBac LESWAS (a), LDSWAS (b), LEEWAS (c), DEWAS (d), LEDWAS (e) and LDDWAS (f) mutant channels for $Na^+$, $Li^+$, $K^+$, $Rb^+$ and $Cs^+$ (as labelled) were determined by normalising peak current magnitudes recorded in $Na^+$ bath solution from the same cell prior to replacement of extracellular $Na^+$ for test cation. Averages (±SEM) are from at least 5 cells.

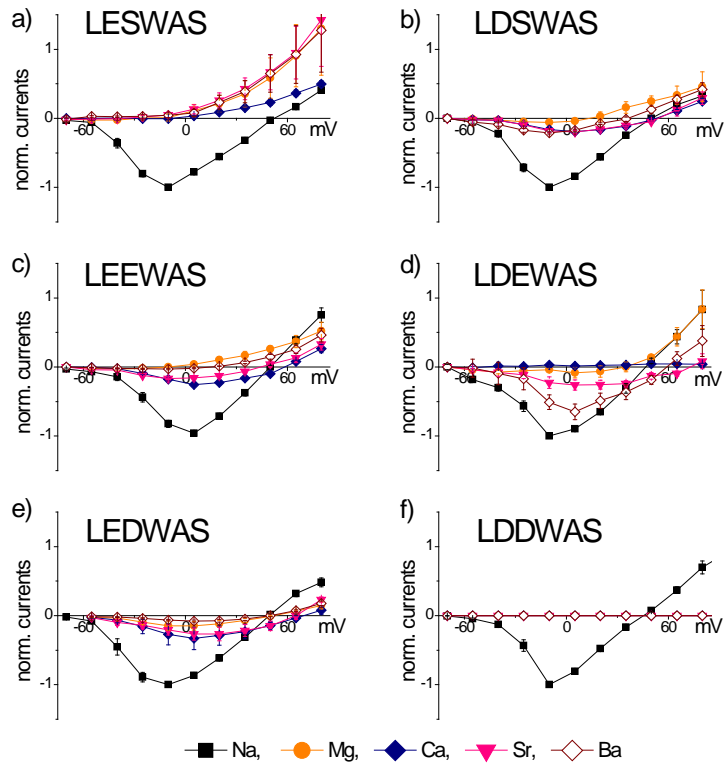

**Supplemental Figure 4.** Divalent cation permeability. Mean peak current voltage relationships for wild type NaChBac LESWAS (a), LDSWAS (b), LEEWAS (c), LDEWAS (d), LEDWAS (e) and LDDWAS (f) mutant channels for $Na^+$ (for comparison), $Mg^{2+}$, $Ca^{2+}$, $Sr^{2+}$ and $Ba^{2+}$ (as labelled) were determined by normalising peak current magnitudes recorded in Na+ bath solution from the same cell prior to replacement of extracellular $Na^+$ for test cation. Averages (±SEM) are from at least 5 cells.


\end{document}